# Hydrolytic Degradability, Cell Tolerance and On-Demand Antibacterial Effect of Electrospun Photodynamically Active Fibres

Amy Contreras [1], Michael J. Raxworthy [1,2], Simon Wood [3] and Giuseppe Tronci [3,4,*]

[1] Institute of Medical and Biological Engineering, University of Leeds, Leeds LS2 9JT, UK

[2] Neotherix Ltd., The Hiscox Building, Peasholme Green, York YO1 7PR, UK

[3] School of Dentistry, St. James's University Hospital, University of Leeds, Leeds LS9 7TF, UK

[4] Clothworkers Centre for Textile Materials Innovation for Healthcare, School of Design, University of Leeds, Leeds LS2 9JT, UK

* Correspondence: g.tronci@leeds.ac.uk

## Abstract

Photodynamically active fibres (PAFs) are a novel class of stimulus-sensitive systems capable of triggering antibiotic-free antibacterial effect on-demand when exposed to light. Despite their relevance in infection control, however, the broad clinical applicability of PAFs has not yet been fully realised due to the limited control in fibrous microstructure, cell tolerance and antibacterial activity in the physiologic environment. We addressed this challenge by creating semicrystalline electrospun fibres with varying content of poly[(L-lactide)-co-(glycolide)] (PLGA), poly(ε-caprolactone) (PCL) and methylene blue (MB), whereby the effect of polymer morphology, fibre composition and photosensitiser (PS) uptake on wet state fibre behaviour and functions was studied. The presence of crystalline domains and PS–polymer secondary interactions proved key to accomplishing long-lasting fibrous microstructure, controlled mass loss and controlled MB release profiles (37 °C, pH 7.4, 8 weeks). PAFs with equivalent PLGA:PCL weight ratio successfully promoted attachment and proliferation of L929 cells over a 7-day culture with and without light activation, while triggering up to 2.5 and 4 log reduction in *E. coli* and *S. mutans* viability, respectively. These results support the therapeutic applicability of PAFs for frequently encountered bacterial infections, opening up new opportunities in photodynamic fibrous systems with integrated wound healing and infection control capabilities.

**1. Introduction**

Antibiotic resistance poses a risk to mankind, threatening not just our health and food safety but also our environment [1]. If not treated in a timely manner, bacterial contamination causes infection and detrimentally affects the performance of medical devices implanted in vivo as well as the outcome of surgical therapies. With the growing trends in antibiotic resistance, timely management of infection is therefore at risk, especially in the context of chronic ulcers [2], oral wounds [3] and bone fractures [4], which frequently cause significant pain, death risks and economic burden. Surgical wound infections alone account for one-quarter of all healthcare-associated infections, causing an estimated 2.5-fold increase in hospital stay and additional healthcare costs [5]. There is therefore a critical need for antimicrobial strategies that minimise antibiotic reliance and that treat antibiotic-resistance infections when activated on-demand via safe external stimuli.

Photodynamic therapy (PDT) is a promising antibiotic-free antimicrobial treatment, which is based on the application of light to a photosensitiser (PS) in the presence of molecular oxygen. The light is absorbed by the PS, enabling the transfer of an electron from a ground to an excited state. The consequent reaction of the excited PS molecule with molecular oxygen generates toxic reactive oxygen species (ROS) [6]. Due to the high reactivity of ROS, PDT can be conveniently used to eradicate fungi, viruses [7] and bacteria [8], regardless of resistant strains and with minimal risks of inducing antimicrobial resistance. Consequently, the application of PS and light for PDT should therefore be tailored to accomplish antibacterial photodynamic effect, on the one hand, and cellular viability and low dark cytotoxicity (i,e., in the absence of light), on the other hand.

Variation of PS dosage and light intensity can be exploited to control the extent of photodynamic effect; however, accomplishing selective antimicrobial activity is still a challenge due to the concentration-dependent aggregation of PS molecules [9], which can lead to limited ROS generation [10], water solubility [6,11] and visible light absorption [12]. These limitations restrict the applicability of PDT in the clinical management of infectious diseases [13], at a time when the need for new antimicrobial strategies has never been more critical. Although the synthesis and testing of new PSs to address the above-mentioned challenges has been reported [10,11,14], encapsulation of PSs already in use in the clinic in polymer carriers represents a promising route to minimise PS aggregation, ensure a prolonged therapeutic window and comply with current regulatory framework [15–18]. With this strategy, control of the (i) release profile of PS from the carrier, (ii) PS uptake in both cells and bacteria and (iii) selectivity between the bactericidal effect and sparing effect upon host cells are key to successful performance.

Electrospun meshes are frequently used for the development of wound [19] and oral [20] dressings, whereby the fibre diameter and pore size can be tuned to accommodate drugs and support soft tissue repair and infection control [21–25]. By selecting appropriate fibre-forming building blocks, bespoke molecular configurations [26,27] and drug–polymer nanoscale interactions [28], the electrospinning process can be developed to enhance hydrolytic fibre stability and the retention of both drugs and fibres in wet environment. Aliphatic polyesters, e.g., PLGA, have been used for the fabrication of electrospun biomedical fibres [29,30], as they can be used in FDA-approved devices, e.g., surgical sutures [31]; are biodegradable; and can release degradation products that support wound repair [32,33]. In light of their superior processability, these polymers can be encapsulated with soluble factors during fibre formation, so that factor release can be achieved through either fibre erosion, diffusion or swelling [34].

In this work, we describe the simple and effective fabrication of long-lasting electrospun PAFs that induce bactericidal effect on both Gram-negative *E. coli* and Gram-positive *S. mutans* when irradiated with visible light, while preserving the viability of mammalian cells. Our strategy to deliver on this was to leverage both the photodynamic capability of single PSs and the semicrystalline morphology of fibre-forming aliphatic polyesters, so that PAFs with controlled photodynamic effect, controlled PS release capability and retained wet state fibrous architecture could be successfully accomplished, while avoiding multistep synthetic routes of polymer functionalisation and post-spinning fibre crosslinking. In solution, methylene blue (MB) displayed higher uptake in either aforementioned bacteria, i.e. *E. coli* and *S. mutans*, or L929 mouse fibroblasts with respect to erythrosine (ER), and was therefore selected as a suitable PS for the creation of PAFs. Electrospinning of MB-loaded solutions containing varying PLGA:PCL weight ratios generated submicron fibres with ~100 wt.% encapsulation efficiency and up to ~1.3 MPa of averaged Young's modulus. The hydrolytic degradation and MB release profiles of PAFs were strongly affected by the polymer morphology and chemical composition, so that PAFs could successfully support attachment and proliferation of L929 mouse fibroblasts over a 7-day cell culture and still trigger up to 4 log reduction in bacteria viability. The knowledge gained with these electrospun PAFs could be useful to accomplish photodynamic systems with integrated anti-infection and wound healing functionalities for the management of chronic tissue states.

## 2. Materials and Methods

### 2.1. Materials

Poly[(L-lactide)-co-(glycolide)] (PLGA1090) (PURASORB PLG 1017, $M_n$: 63,000 g·mol$^{-1}$, 10:90 molar ratio of L-lactide and glycolide units) was purchased from Corbion (Gorinchem, The Netherlands). Poly($\varepsilon$-caprolactone) (PCL) ($M_n$: 80,000 g·mol$^{-1}$), MB and ER were supplied by Sigma Aldrich (Gillingham, United Kingdom). Octylphenol Decaethylene Glycol Ether (OPDEGE) and 1,1,1,3,3,3-hexafluoro-2-propanol (HFIP) were purchased from Alfa Aesar (Heysham, United Kingdom) and Fluorochem Ltd. (Hadfield, United Kingdom), respectively.

Phosphate-buffered saline (PBS) solution, Trypsin and completed Minimum Essential Media Eagle–alpha modification (α-MEM) (1 *w/v*% L-Glutamine (Gln), 10 vol.% Foetal Bovine Serum (FBS) and 1 *w/v*% Penicillin/Streptomycin (P/S)) were purchased from Lonza (Slough, United Kingdom). Brain Heart Infusion (BHI) broth and Columbia Blood Agar Base (CB) were supplied by Oxoid. Horse Blood Oxalated was purchased from Thermo Fisher Scientific (Altrincham, United Kingdom).

### 2.2. Electrospinning of Photodynamically Active Fibres (PAFs)

Polymers were stirred in HFIP for 48 h in dark, as per our previous protocol [16]. PLGA and PCL were dissolved in varied weight ratios (PLGA:PCL = 20:80; 50:50;

80:20) with an overall polymer concentration of 6 wt.%. The electrospinning polymer solutions were supplemented with either 2.2 mM, 1.1 mM or 0.22 mM of MB, and transferred into a 10 mL plastic syringe with an 18-gauge blunt-ended needle. The syringe was connected to a syringe pump and a pump rate of 0.03 mL·min$^{-1}$ was used. A cylindrical grounded mandrel (height = 125 mm, diameter = 75 mm) was coated with aluminium foil at 100 mm distance away from the needle tip and rotated at 30 rpm. Fibres were electrospun in dark with an electrostatic voltage of 16 kV and dried under reduced pressure for 72 h. Table 1 summarises the nomenclature and chemical composition of electrospun samples in the form of PAFs and MB-free fibre controls. Samples of PAFs are denoted as *MBXX-PLGAY-CLZ*, whereby *XX* identifies the concentration of MB in the electrospinning solution (*20*: 2.2 mM, *10*: 1.1 mM; *2*: 0.2 mM). *PLGA-CL* indicates the blend of PLGA and PCL, while *Y* and *Z* refer to the weight fraction of respective polymers (i.e., 20, 50 or 80). MB-free controls are coded as *PLGAY-CLZ*.

**Table 1.** Nomenclature and chemical composition of electrospun samples. [a] Concentrations of PLGA, PCL and MB in the fibre-forming electrospinning solution.

| Sample ID | PLGA (wt.%) [a] | PCL (wt.%) [a] | MB (mM) [a] |
|---|---|---|---|
| MB20-PLGA20-CL80 | 20 | 80 | 2.2 |
| MB20-PLGA50-CL50 | 50 | 50 | 2.2 |
| MB20-PLGA80-CL20 | 80 | 20 | 2.2 |
| MB2-PLGA50-CL50 | 50 | 50 | 0.2 |
| MB10-PLGA50-CL50 | 50 | 50 | 1.1 |
| PLGA20-CL80 | 20 | 80 | 0 |
| PLGA50-CL50 | 50 | 50 | 0 |
| PLGA80-CL20 | 80 | 20 | 0 |

*2.3. Scanning Electron Microscopy (SEM)*

Gold-coated dry fibres were imaged with a Hitachi Scanning Electron Microscope (Maidenhead, United Kingdom) in variable pressure low vacuum mode

(270 Pa). Five randomly selected locations were analysed for each sample, and ten fibre diameters measured for each location.

*2.4. Differential Scanning Calorimetry*

Differential Scanning Calorimetry (DSC) temperature scans were recorded (Q100 – TA Instruments, New Castle, DE, USA) on dry electrospun samples (~10 mg). The first DSC run was carried out to remove thermal history and conducted from 20 °C to 250 °C at a heating rate of 10 °C·min$^{-1}$. Subsequently, the sample was cooled down at a cooling rate of 5 °C·min$^{-1}$ and the previous heating cycle was applied again to take the measurement. The DSC thermograms were plotted with endothermic transitions looking downwards.

*2.5. Hydrolytic Degradation Tests*

Dry electrospun samples (n = 3) were cut into 1 cm$^2$ squares and weighed prior to 8-week incubation in PBS (37 °C, pH 7.4) with enough volume (5–10 mL) to maintain the solution pH at ~7.4 over time. At selected time points, samples were collected, rinsed with distilled water and dried in a vacuum desiccator for 1 week. The mass loss was calculated via Equation (1):

$$Mass\ loss = \frac{m_t - m_d}{m_d} \times 100 \qquad (1)$$

whereby $m_t$ and $m_d$ represent the dry weights of both the sample collected at the selected time point $t$ and the fresh electrospun sample, respectively.

*2.6. Methylene Blue (MB) Release Tests*

Dry samples (~20 mg) were incubated in PBS (37 °C, 4 weeks) with enough volume to maintain the pH at ~7.4. At selected time points, 100 μL of the degrading medium was collected, measured by UV–Vis spectrophotometry, and added back to the sample following the measurement. The collected solutions (100 µL) were analysed on a microplate reader to record peak absorbance at 610 nm, corresponding to the wavelength of MB maximum absorbance. A calibration curve ($R^2$ > 0.99) was built by measuring the absorbance of varying MB-supplemented PBS solutions to convert absorbance into MB concentration values.

*2.7. Photosensitiser Uptake Study*

2.7.1. Uptake Study with L929 Cells

L929 cells were seeded on to a 96-well plate at a density of 5 × $10^3$ cells·$mL^{-1}$ (100 µL) and incubated for 24 h to allow for cell attachment. Resulting confluent monolayers were exposed to 2-h incubation with four MB-supplemented (0.2, 2, 20 and 200 µg·$mL^{-1}$) PBS solutions. PBS solutions with ER (0.2–200 µg·$mL^{-1}$) were also tested, whereby ER was selected as an additional PS. Control wells contained either the cell-free PS-supplemented PBS solutions or the cell-supplemented PBS solution. Following 2-h incubation, the solutions were removed, and each well was washed with fresh PBS, prior to application of a 10 vol.% OPDEGE solution in PBS. The well plate was shaken for 5 min and incubated for 25 min. Absorbance measurements were recorded at either 610 nm (for MB groups) or 530 nm (for ER), whereby the background interference from the absorption values of wells containing only solubilised cells was subtracted. A calibration curve with PS-supplemented PBS solutions was built to calculate the number of moles of PS taken up by the cells.

2.7.2. Uptake Study with Bacteria

*S. mutans* Ingbritt (Gram-positive coccus bacteria) and *E. coli* 11,954 (Gram-negative rod-shaped bacteria) were selected as bacterial strains for this study. An overnight culture of each bacterial strain was prepared, and 1 mL of overnight culture added to 9 mL of fresh BHI. Following incubation to mid-log phase (3.5 h for *S. mutans* and 1.5 h for *E. coli*), the absorbance of the suspension was measured at 600 nm ($OD_{600}$) and the suspension diluted in fresh BHI to achieve a final concentration of $10^8$ CFU·mL$^{-1}$. The planktonic bacteria suspension was washed once in PBS to remove culture media contaminants and then resuspended in PBS solutions supplemented with either MB or ER (20, 200 µg·mL$^{-1}$), prior to transfer into a 96-well plate (100 µL per plate) and 2-h incubation. Control wells contained either the PS solution with no bacteria or fresh PBS with bacteria. Following incubation, the bacteria were washed again in PBS solution to remove extracellular PS. The PBS solution was then replaced with a 10 vol.% solution of OPDEGE in PBS and well plates were shaken for 5 min and left to incubate for 25 min to ensure complete bacterial lysis, prior to absorbance measurements, as reported in Section 2.7.1.

## 2.8. Cytotoxicity Tests

Dry electrospun samples were cut into squares (~20 mg) and exposed to 15 min of UV light on each side. Resulting samples were then employed for both extract and contract cytotoxicity tests, as reported below.

### 2.8.1. Extract Cytotoxicity Tests

Extract sample solutions were prepared by incubating previously accomplished UV-disinfected samples in 5 mL of PBS at 37 °C for 0, 2 and 24 h. In parallel, L929 murine fibroblasts were seeded onto an opaque walled 96-well plate at a density of $5 \times 10^3$ cells per well with 100 µL of complete α-MEM and incubated for 24 h (37 °C, 5% $CO_2$) to allow for cell attachment. The cell culture medium was then removed and each well washed with fresh PBS, prior to addition of the extract solutions.

Control wells contained fresh PBS with cells (negative control) and 10 vol.% OPDEGE solution in PBS (positive control). Following a 10-min incubation, the well plates were either exposed to light for 60 min or wrapped with aluminium foil and exposed to light for the same period of time (dark control). Following light treatment, an ATP luminescence assay (Promega) was carried out to quantify cell viability, according to manufacturer instructions. Briefly, 100 µL of CellTiter-Glo® 2.0 solution was added to each well, and the well plate covered with aluminium foil and shaken at 200 RPM for 5 min, prior to 25-min equilibration at room temperature. On each plate, luminosity readings of each tested sample group were compared to the averaged luminosity of the negative dark control (fresh PBS with cells and dark light exposure) to determine the percentage of killed cells, according to Equation (2):

$$Cell\ killed\ (\%) = \frac{Lum_{test}}{Lum_{control}} \times 100 \qquad (2)$$

whereby $Lum_{test}$ and $Lum_{control}$ represent the luminosity of the test and control wells, respectively.

2.8.2. Contact Cytotoxicity Tests

UV-disinfected samples were individually placed in a 48-well plate and incubated (37 °C, 5% $CO_2$) for 60 min with 0.5 mL of sterile supplemented α-MEM to allow for sample equilibration. Confluent L929 cells were seeded onto the centre of each sample with a seeding density of 5 × $10^3$ cells per sample, prior to 60-min light exposure. Dark controls were placed in the 48-well plate, and the plate wrapped around aluminium foil and exposed to 60-min light irradiation. Following this, samples were incubated (37 °C, 5% $CO_2$) for either 24 h or 7 days (with media changed every other day). Upon completion of the cell culture, samples were collected, washed twice with fresh PBS and placed in formalin for 24 h. The samples were dehydrated in a series of increasing ethanol concentrations (20–100 vol.%) in water, prior to SEM (Section 2.3) to visualise cell attachment and morphology.

*2.9. Antimicrobial Photodynamic Tests*

Overnight *S. mutans* and *E. coli* cultures were prepared by inoculating a single colony of bacteria in 20 mL of BHI overnight. The following day, 1 mL of this broth was added to 9 mL of fresh BHI and the solution incubated until mid-log phase (3.5 h for *S. mutans* and 1.5 h for *E. coli*). The number of bacteria was estimated by recording $OD_{600}$ and the concentration adjusted to a final concentration of $5 \times 10^8$ $CFU \cdot mL^{-1}$. Resulting bacteria cultures were then employed for either extract or contact photodynamic tests as reported below.

2.9.1. Extract Photodynamic Tests

Extract sample solutions were prepared as reported in Section 2.8.1, i.e., by incubating UV-disinfected samples in 5 mL of PBS at 37 °C for 0, 2 and 24 h. Bacteria were washed once in PBS before being resuspended in 10% extract solutions. Fresh PBS and 10 vol.% OPDEGE solution in PBS were used as negative and positive control, respectively. Bacteria suspensions were then seeded (100 µL) onto an opaque walled 96-well plate and immediately exposed to light for 60 min, or wrapped with aluminium foil and exposed to light for the same period of time (dark control). Following light treatment, the solutions were serially diluted and 100 µL aliquots were spread onto fresh agar plates. The plates were then incubated at 37 °C for 24 h. The following day, manual counting of colonies was performed and the number of CFU for each solution was calculated and compared to the initial inoculation. The experiment was performed in triplicate.

2.9.2. Contact Photodynamic Tests

Lawns of bacteria were spread by using a sterile swab dipped in bacterial broth and streaking across a fresh agar plate. Sterile tweezers were used to place UV-disinfected electrospun discs (Ø 10 mm, n = 3) onto the agar plate inoculated with bacteria in triplicate. The plates were incubated for 60 min to allow for MB release, before being exposed to light for 30 or 60 min (or wrapped in foil as the dark control). Agar plates were then incubated at 37 °C overnight to allow for bacterial growth. The following day, images were taken of each plate. Zones of inhibition were calculated using the straight-line tool on ImageJ software, to count the number of pixels in a known section of the image to produce a scale, and then to measure the zone size for each image.

*2.10. Statistical Analysis*

Significant differences were evaluated using an unpaired Student's t-test ($p < 0.05$). All data were collected in triplicate and presented as mean ± standard deviation.

## 3. Results and Discussion

Electrospinning has recently been shown to generate PAFs with significant antibacterial photodynamic effect against multiple bacteria [16–18,22,33,35]. Despite these initial advances, however, the biocompatibility of PAFs in both inert and antimicrobial states has only partially been addressed [24,33,35], which is critical to enable applicability in medical devices. Furthermore, electrospun fibres often suffer from instability and macroscopic shrinking in aqueous environments due to water-induced fibre swelling and merging [16,27,35], which can lead to uncontrolled release of PS and fast hydrolytic degradation [34]. To address these challenges, we aimed to build the structure–function relations of PAFs, whereby the effect of fibre-forming polymers, PS type and dosage on fibre micro- and macroscale was studied. Due to the high excess of glycolide units, it was hypothesised that electrospinning of PLGA1090 could enable the development of fibre-stabilising crystalline domains and secondary interactions between the fibre-forming polymer and the PS; while the presence of the PCL phase was expected to generate fibres with decreased hydrolytic degradability [36,37]. In the following, uptake and photodynamic studies were first carried out with MB- and ER-supplemented solutions to explore the response of L929 cells, *E. coli* and *S. mutans* to varying PS concentrations. *E. coli* and *S. mutans* were selected as Gram-negative and Gram-positive bacteria models, given their occurrence in chronic wounds [38] and periodontal diseases [39], respectively. These preliminary investigations informed the selection of the PS for fibre encapsulation, so that PAFs with varied chemical composition (Table 1) were characterised with respect to their hydrolytic stability, release capability, bactericidal effect and cytotoxicity in dark and following light exposure.

*3.1. PS Uptake of Mammalian and Bacterial Cells*

To assess the effectiveness of the PSs in vitro, uptake and photodynamic tests were carried out by exposing L929 cells, *E. coli* and *S. mutans* to a 2-h incubation with either MB- or ER-supplemented PBS solutions with varying PS concentration (0.2–200 µg·mL$^{-1}$). Both mammalian and bacterial cells displayed detectable levels of PS uptake only following incubation with increased PS concentrations (20–200 µg·mL$^{-1}$, 100 µL), whereby significantly increased levels were observed with MB-supplemented solutions (Figure 1). At a MB concentration of 20 µg·mL$^{-1}$ (i.e., 6.25 nmoles MB in 100 µL), L929 mammalian cells displayed the highest uptake of

MB (2.3 nmoles, ~37 mol.%), followed by *S. mutans* (1.9 nmoles, ~30 mol.%) and *E. coli* (1.5 nmoles, ~24 mol.%) (Figure 1A). Incubation with more concentrated solutions (200 µg·mL$^{-1}$, i.e., 62.5 nmoles MB per 100 µL) resulted in increased MB uptake values, although a different trend was observed within the three groups, whereby L929 cells revealed the lowest value of MB content (9 nmoles, ~14 mol.%), followed by *E. coli* (10.8 nmoles, ~17 mol.%) and *S. mutans* (17 nmoles, ~27 mol.%). Similar levels of MB uptake were observed in HeLa epithelial cells [40], whilst higher values were recorded when *E. coli* were exposed to solutions with increased (~50-fold) MB concentration [41]. Compared to MB, cells incubated with ER-supplemented solutions generally displayed lower PS uptake, with the highest value (2.7 nmoles, ~12 mol.%) recorded in L929 cells treated with the more concentrated (200 µg·mL$^{-1}$, i.e., 22.7 nmoles in 100 µL) ER solution (Figure 1B).

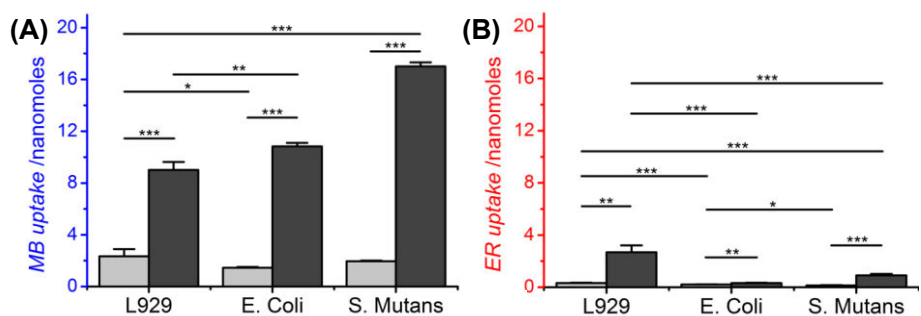

**Figure 1.** Uptake of MB (**A**) and ER (**B**) in L929 cells, *E. Coli* and *S. Mutans* following 2-h incubation in 100 µL of PS-supplemented PBS solution. (■): 20 µg·mL$^{-1}$ PS (i.e., 6.25 nmoles MB, 2.27 nmoles ER, 100 µL); (■): 200 µg·mL$^{-1}$ PS (i.e., 62.5 nmoles MB, 22.7 nmoles ER, 100 µL).* $p < 0.05$, ** $p < 0.01$, *** $p < 0.001$ (t-test, n = 3).

The increased cellular uptake observed with MB- with respect to ER-supplemented solutions can be attributed to the different electrostatic charge and decreased molecular weight of the phenothiazine, compared to the xanthene, dye. The cationic charge of MB has been shown to mediate electrostatic interactions and binding with the bacterial wall and cellular membrane [6,9,42], while the decreased molecular weight facilitates passive diffusion of the PS in the cell, ultimately leading to increased PS uptake [43]. On the other hand, ER displays a dianionic configuration in the neutral pH of the PBS solution [44], whereby the presence of the two negative charges decreases the interaction of ER with the cell membrane, therefore hindering the uptake by bacteria.

### 3.2. PS Impact on L929 Cells, E. coli and S. Mutans

Following the uptake tests, the impact of PS solutions on cellular and bacterial activity was investigated via an ATP luminescence assay, in line with its employability to assess the metabolic activity of cells [45] and bacteria [46], and its minimal interference from chemical compounds and selected light source. ATP measurements were firstly carried out in the absence of light to assess the dark toxicity of both PSs. Low concentration of either MB or ER ([PS] ≤ 2 µg·mL$^{-1}$) proved to marginally affect L929 cells, whereby only ≤ 20% viability reduction was measured with respect to the PBS negative control (Figure 2A).

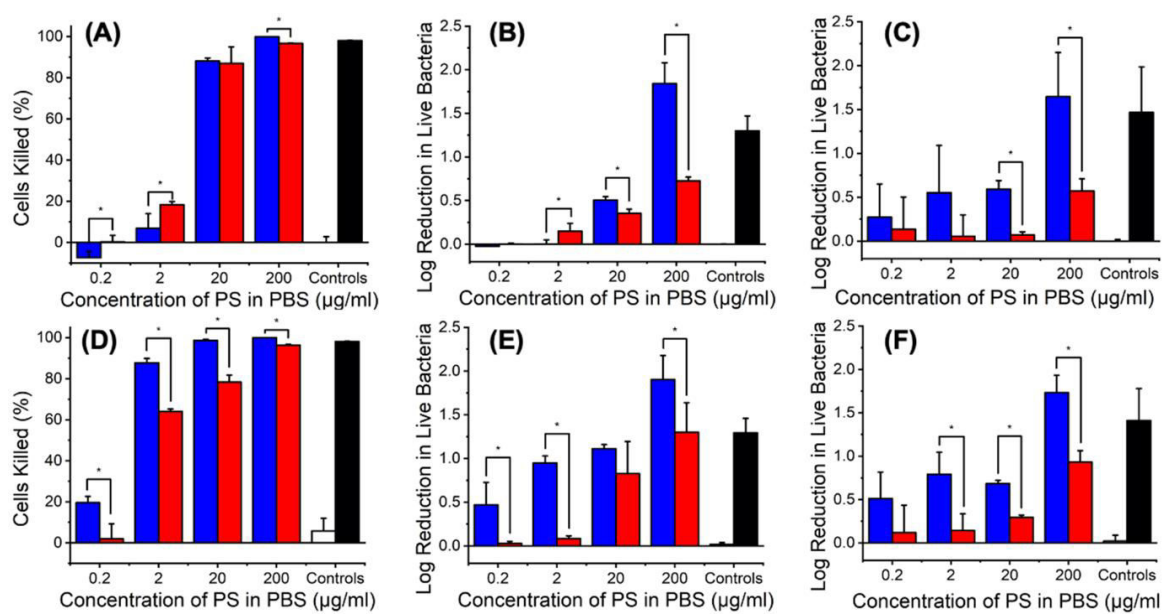

**Figure 2.** ATP toxicity assay on L929 cells (**A,D**), *E. coli* (**B,E**) and *S. mutans* (**C,F**) when treated with PS-supplemented PBS solutions. (**A–C**) 60-min dark incubation. (**D–F**) 60-min light exposure. (■): MB-supplemented solutions. (■): ER-supplemented solutions. (□): PBS negative control. (■): OPDEGE positive control. * $p < 0.05$ (one-way ANOVA with Tukey HSD test, n = 3).

Although minimal effects were also observed with *E. coli* (≤ 0.15 log reduction, < 29%) in the same concentration range ([PS] ≤ 2 µg·mL$^{-1}$, Figure 2B), significant dark toxicity was measured in the same bacteria following exposure to increased concentration ([PS] ≥ 20 µg·mL$^{-1}$) of either MB (0.50–1.84 log reduction, 68–99%) or ER (0.35–0.72 log reduction, 55–81%). Thus, a concentration-dependent toxic response was observed, whereby MB appeared to induce increased dark toxicity with respect to ER, in line with previous PS uptake trends (Figure 1). In comparison to *E. coli*, *S. mutans* exhibited higher dark toxic response (≥ 0.27 log reduction, ≥ 46%) even at low concentration of MB ([MB] ≤ 2 µg·mL$^{-1}$), while lower effects (0.06–0.57 log reduction, 13–73%) were observed with ER-supplemented solutions regardless of the ER concentration (Figure 2C). The decreased dark tolerability of *S. mutans* is supported by the lack of the outer membrane with respect to *E. coli*, as well as by its decreased uptake capability (Figure 1).

When the same ATP assay was carried out following 60-min light exposure (120 mW·cm$^{-2}$, 215 J·cm$^{-2}$, see Supplementary Information), an obviously higher reduction of L929 cells (≥ 60%) occurred at [PS] ≤ 2 µg·mL$^{-1}$ compared to respective dark measurements (Figure 2D), whereby MB still displayed the highest cytotoxic effect. In comparison with L929 cells, ≤ 0.9 log reduction (≤ 87%) of *E. coli* viability was observed with either [MB] ≤ 2 µg·mL$^{-1}$ or [ER] ≤ 20 µg·mL$^{-1}$, respectively (Figure 2E), while at least 0.83 log reduction (> 85%) was recorded at increased concentrations. Ultimately, only a marginal increase in viability reduction was measured on *S. mutans* following light activation with respect to the dark measurements (Figure 2F), whereby ≤ 1.73 (98%) and ≤ 0.93 (88%) log reductions were measured with MB and ER, respectively.

Overall, these results confirmed the photodynamic activity of both PSs against selected bacteria, in line with previous reports [47]. The photodynamic effect of selected ER-supplemented solutions proved to be lower than previous antibacterial results with *S. mutans* (≥ 2 log reduction) [44], a result that is attributed to the decreased concentration of ER (0.23–230 μM) used in this study. The above-mentioned results therefore demonstrated the superior photodynamic activity of MB across L9292 cells, *E. coli* and *S. mutans* across all concentrations, so that MB was selected for the fabrication of PAFs. An MB concentration below 2 μg·mL$^{-1}$ (< 63 μM) was expected to show increased tolerability of L929 cells in both dark and photodynamic conditions, while still ensuring ≤ 0.8 antibacterial effect following activation with light. In the following, the polymer morphology in and degradability of PAFs as well as MB dosage will be investigated aiming to achieve fibres with controlled release capability of PS in near physiologic environments. These molecular characteristics will then be varied aiming to accomplish long-lasting PAFs with independently controlled cellular tolerance and on- demand photodynamic effect.

*3.3. Microstructure of Electrospun PAFs*

Electrospun samples with varying PLGA:PCL ratios were successfully accomplished with ~100 wt.% encapsulation efficiency (see Equation (S1), Supplementary Information), confirming full transfer of MB from the electrospinning solution to the fibres. SEM revealed homogeneous and defect-free fibre morphology regardless of the selected PLGA:PCL weight ratio (Figure 3A–C). The fibre diameter was measured in the range of 0.44 ± 0.18–0.83 ± 0.22 μm and proved to be significantly smaller than the one measured in the MB-free fibre controls (Ø = 1.15 ± 0.54–1.68 ± 0.52 μm, *p* < 0.001) (Table 2, Figure S1A, Supplementary Information). Similar values have been reported in similar electrospun polyesters [48,49], while the decreased fibre size recorded in MB-encapsulated samples is attributed to the electrostatic repulsion of fibre-forming electrospinning jets containing the positively charged molecule [16,26].

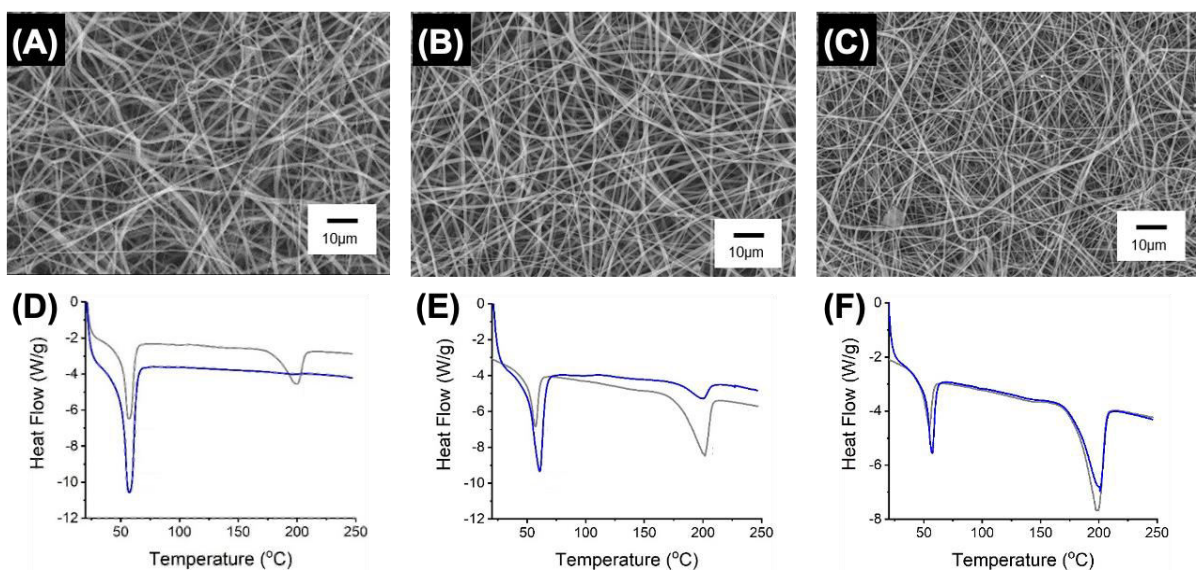

**Figure 3.** SEM and DSC analyses of electrospun fibres MB20-PLGA20-CL80 (**A**,**D**), MB20-PLGA50- CL50 (**B**,**E**) and MB20-PLGA80-CL20 (**C**,**F**). Grey DSC curves refer to the MB-free fibre controls.

Other than the effect of MB, sample MB20-PLGA50-CL50 prepared with equivalent PLGA:PCL weight ratio displayed significantly increased fibre diameter (Ø = 0.83 ± 0.22 µm) with respect to fibres containing either increased (Ø = 0.54 ± 0.18 µm, MB20-PLGA80-CL20, $p < 0.01$) or decreased (Ø = 0.44 ± 0.17 µm, MB20-PLGA20-CL80, $p < 0.001$) PLGA content (Table 2, Figure S1A, Supplementary Information). This observation was hypothesised to be attributed to the poor miscibility of PLGA and PCL [50], which was expected to be decreased when equivalent weight fractions of both polymers were applied. The poor miscibility of selected fibre-forming polymers was supported by DSC measurements (Figure 3D–F).

Both samples MB20-PLGA50-CL50 and MB20-PLGA80- CL20, and their respective MB-free controls, revealed two distinct endothermic transitions, corresponding to the melting of PCL ($T_{m1}$ ~60 °C) and PLGA ($T_{m2}$ ~200 °C) related crystalline domains (Table 2). Furthermore, the overall melting enthalpy ($\Delta H$ = 79 J·g$^{-1}$) recorded with sample MB20-PLGA50-CL50 (Figure 3E, Table 2) was higher than the one measured with the other two MB-encapsulated samples, thereby supporting the above-mentioned variation in fibre diameters.

Unlike fibres made with increased PLGA content, however, a different thermal behaviour was observed in PCL-rich PAFs MB20-PLGA20-CL80 and respective MB-free variant. While the former sample revealed only the PCL-related melting transition, both PCL- and PLGA-related melting transitions were detected in the fibre control (Figure 3D). The most likely explanation for this observation is that the MB molecules encapsulated in the fibre are more compatible with the PLGA than with the PCL phase. The excess of carbonyl groups in the repeat unit of the former compared to the latter polymer may promote increased electrostatic interactions with MB, so that PLGA crystallisation is hindered in MB-encapsulated fibres, similarly to other polyester fibres loaded with low molecular weight additives [37,49]. This observation is supported by the constantly decreased PLGA-related melting enthalpy ($\Delta H_{m2}$) measured in PAFs compared to the corresponding fibre controls (Table 2).

**Table 2.** Fibre diameter (Ø), melting transition temperatures $T_m$ and melting enthalpies ($\Delta H_m$) recorded in electrospun samples via SEM and DSC, respectively. N.o.: not observed. N.a.: not available.

| Sample ID | Ø (µm) | $T_{m1}$ (°C) | $\Delta H_{m1}$ (J·g$^{-1}$) | $T_{m2}$ (°C) | $\Delta H_{m2}$ (J·g$^{-1}$) |
|---|---|---|---|---|---|
| MB20-PLGA20-CL80 | 0.44 ± 0.17 | 57 | 72 | n.o. | n.o. |
| MB20-PLGA50-CL50 | 0.83 ± 0.22 | 61 | 59 | 200 | 20 |
| MB20-PLGA80-CL20 | 0.54 ± 0.18 | 61 | 16 | 201 | 47 |
| MB2-PLGA50-CL50 | 0.76 ± 0.23 | 58 | 32 | 201 | 30 |
| MB10-PLGA50-CL50 | 0.83 ± 0.29 | n.a. | n.a. | n.a. | n.a. |
| PLGA20-CL80 | 1.68 ± 0.52 | 57 | 33 | 200 | 37 |
| PLGA50-CL50 | 1.48 ± 0.56 | 61 | 22 | 202 | 44 |
| PLGA80-CL20 | 1.15 ± 0.54 | 57 | 10 | 197 | 58 |

In order to further elucidate the effect of either MB encapsulation or PLGA:PCL weight ratio on the macroscopic behaviour of respective fibres, tensile tests were carried out (Figure 4). Significantly decreased strain at break ($\varepsilon_b$ = 119 ± 122–199 ± 2%) and increased Young's modulus ($E$ = 0.49 ± 0.05–1.32 ± 0.20 MPa) were recorded in all PAFs compared to respective fibre controls ($\varepsilon_b$ = 278 ± 25–378 ± 48%; $E$ = 0.076 ± 0.002–0.212 ± 0.019 MPa; $p < 0.05$), as well as within PAFs with increased PLGA content (Figure 4A,C). These variations in tensile properties support the development of PLGA-MB secondary interactions at the molecular scale of the fibre, in agreement with previous DSC measurements (Figure 3D–E, Table 2). Other than that, MB-free fibres made with increased PLGA content revealed a gradual increase of stress at break (Figure 4B) and Young's modulus (Figure 4C), a result that is in agreement with the inherently higher mechanical properties of PLGA with respect to PCL fibres and the non-miscibility of the two polymers [49,51–53].

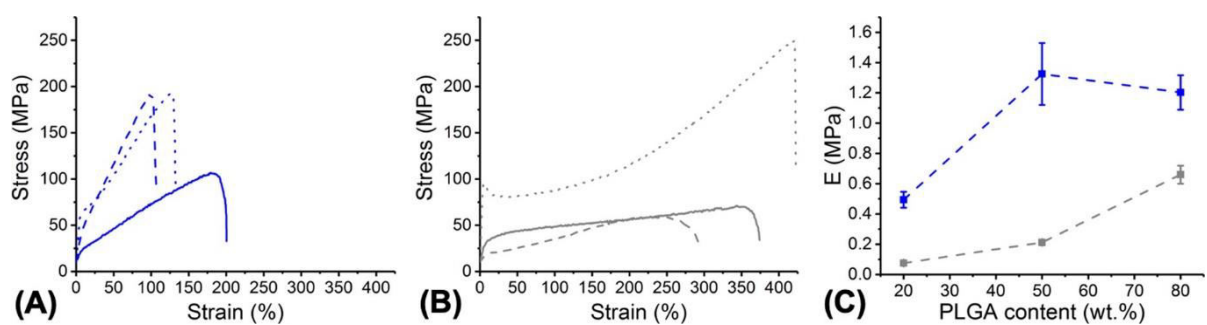

**Figure 4.** Typical tensile stress–strain curves of PAFs (**A**) and MB-free fibre controls (**B**). (—): MB20-PLGA20-CL80; (---): MB20-PLGA50-CL50; (···): MB20-PLGA80-CL20. (—): PLGA20-CL80; (---): PLGA50-CL50; (···): PLGA80-CL20. (**C**) Variation of Young's modulus (E) in PAFs (■) and MB-free controls (■) made with varied PLGA content. Lines are guidelines to the eye.

Electrospinning of solutions with decreased MB concentration ([MB] = 0.22 and 1.1 mM) was also carried out, given the effect of MB dosage on cell viability (Figure 2) and aiming to realise PAFs with varied antimicrobial photodynamic effect and preserved cell tolerability. To deliver on this, the formulation PLGA50-CL50 was selected due to the increased tensile modulus (Figure 4) and crystallinity of respective electrospun materials (Figure 3). A regular fibre morphology was observed in both samples, MB10-PLGA50-CL50 (Ø = 0.83 ± 0.29 µm) and MB2-PLGA50-CL50 (Ø = 0.76 ± 0.23 µm) (Figure 5), whereby a decreasing trend in fibre diameter was observed in PAFs encapsulated with increased content of MB (Figure S1B, Supplementary Information), in agreement with previous results (Figure 3). Nevertheless, no significant difference was measured with respect to sample MB20-PLGA50-CL50 (Figure S1B, Supplementary Information), indicating that the selected range of MB dosage ([MB] = 0.2–2 mM) did not affect the fibre morphology.

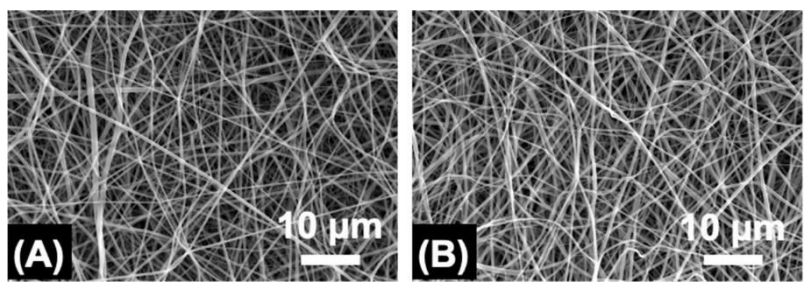

Figure 5. SEM of electrospun fibres MB2-PLGA50-CL50 (**A**) and MB10-PLGA50-CL50 (**B**).

Other than SEM, the DSC thermogram of sample MB2-PLGA50-CL50 still displayed two melting transitions of PLGA and PCL domains, although a lower overall melting enthalpy ($\Delta H$ = 62 J·g$^{-1}$) was measured with respect to fibres MB20-PLGA50-CL50 ($\Delta H$ = 79 J·g$^{-1}$) (Table 2). While this result confirms the non-miscibility of the PLGA and PCL domains, it also suggests that increased PLGA-MB secondary interactions are developed when decreased content of MB is encapsulated in the fibre.

*3.4. Fibre Behaviour in Physiological Environment*

Additionally to microstructure and thermal analysis, the attention moved to the characterisation of PAFs in an aqueous environment, aiming to accomplish fibres with controlled degradability, retained fibrous architecture and sustained release of MB. The equilibrium water uptake of both MB- encapsulated and MB-free samples was initially measured, as the first event in the hydrolytic degradation process of aliphatic polyesters [36,54,55]. No detectable macroscopic changes (e.g., shrinking) were observed in the retrieved 24-h incubated fibres, while a significantly increased water uptake was measured in PAFs ($WU$ = 301 ± 2–348 ± 25 wt.%) with respect to MB-free controls ($WU$ = 62 ± 6–250 ± 19 wt.%, $p < 0.05$) (Figure 6A). MB-free fibres made with the highest PLGA content also displayed a significant increase in water uptake with respect to the other two sample groups, while a counterintuitive opposite trend was observed with respective MB- encapsulated samples (Figure 6A).

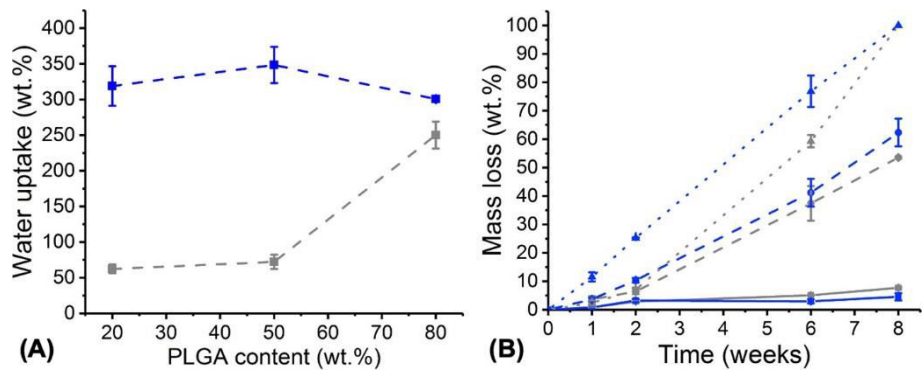

**Figure 6.** (**A**) Water uptake of electrospun samples presenting varied PLGA content with (■) and without (■) MB (2.2 mM). (**B**) Mass loss of electrospun samples during 8-week incubation in PBS (pH 7.4, 37 °C). (■): MB20-PLGA20-CL80. (■): PLGA20-CL80. (●): MB20-PLGA50-CL50. (●): PLGA50-CL50. (▲): MB20-. PLGA80-CL20. (▲): PLGA80-CL20. Results are reported as Mean ± SD (n = 3). Lines are guidelines to the eye.

Polymer crystallinity is known to play a critical role on the diffusion of water in aliphatic polyesters [36,54,55], as the presence of ordered domains limits water diffusion, water uptake and volumetric swelling of fibres. The semicrystalline morphology of both PAFs and fibre controls therefore explains the dimensional stability of these samples following incubation with water, in contrast to the prompt macroscopic shrinking observed when amorphous fibrous architectures were tested [16,27,35]. The increased values of water uptake measured in PAFs compared to the corresponding fibre controls therefore reflects aforementioned considerations, as decreased melting enthalpies, suggesting lower crystallinity [56], were constantly measured in the former compared to the latter groups (Table 2).

Together with the polymer morphology, the hydrophilicity and secondary interaction capability of the fibre-forming polymer are also key in explaining the lowest and highest water uptake of samples MB-PLGA80-CL20 and PLGA80-CL20 among each sample group, respectively. PLGA is known to be more hydrophilic compared to both PLA and PCL [36,55], given the absence of methyl groups and reduced number of methylene bridges in respective repeat unit. Consequently, the higher water uptake measured in samples PLGA80-CL20 compared to both samples PLGA50-CL50 and PLGA20- CL80 is attributed to the excess of PLGA content and respective increase in fibre hydrophilicity in the former with respect to the two latter groups. The presence of PLGA is also likely to explain the counterintuitively lower water uptake of sample MB20-PLGA80-CL20 with respect to the other two samples of PAF (Figure 6A), due to the presence of PLGA-MB secondary interactions (Figure 3D).

Following the water uptake measurements, an 8-week hydrolytic degradation study was carried out, whereby the sample mass loss (Figure 6B) and microstructure (Figure 7) were assessed at specific time points.

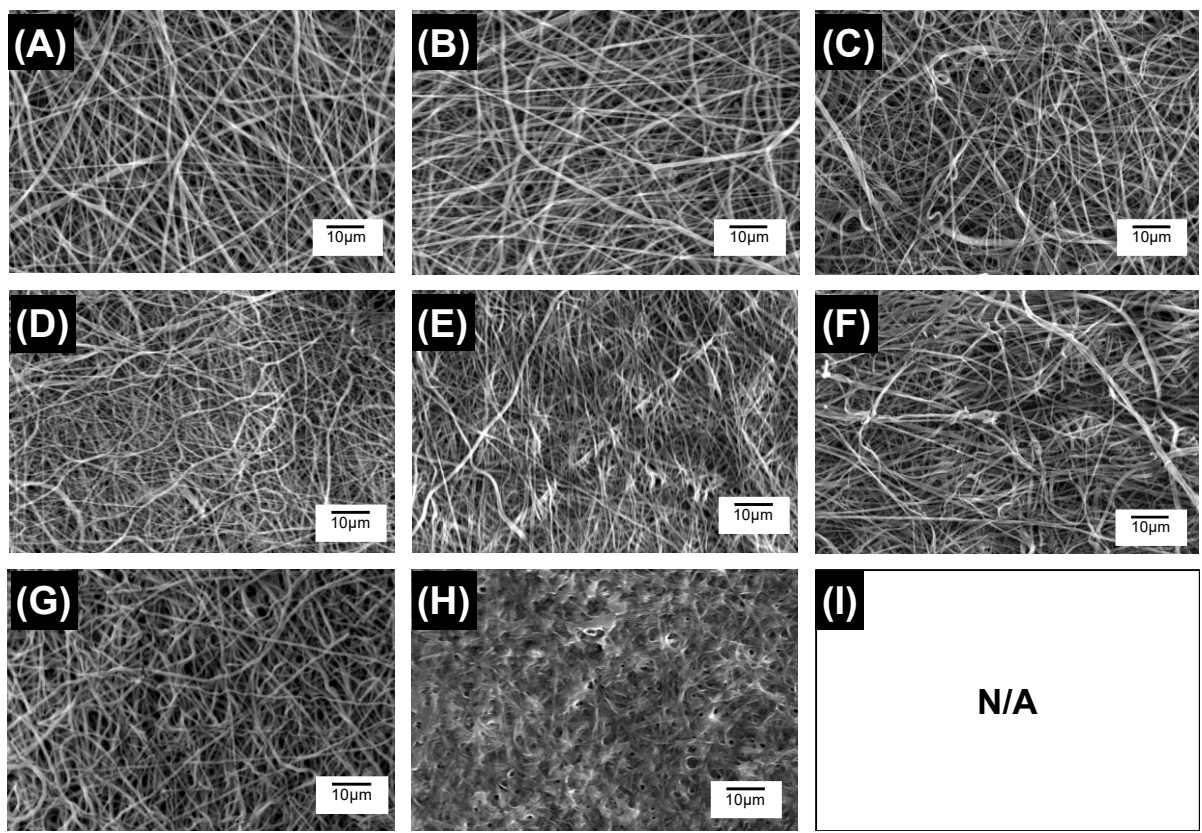

**Figure 7.** SEM of MB-encapsulated fibres during 8-week hydrolytic degradation (37 °C, pH 7.4). (**A**–**C**) MB20-PLGA20-CL80 at week 2 (A), 6 (B) and 8 (C). (**D**–**F**) MB20-PLGA50-CL50 at week 2 (D), 6 (E) and 8 (F). (**G**–**I**) MB20-PLGA80-CL20 at week 2 (G), 6 (H) and 8 (I) (completely degraded sample).

Regardless of the presence of MB, samples with increased PLGA content degraded more rapidly than samples with decreased PLGA content, while both samples

MB20-PLGA20-PCL80 and PLGA20-PCL80 exhibited the lowest change in weight (Figure 6B). These trends in mass loss were also supported by the changes in fibrous architectures observed in retrieved samples over the course of hydrolytic degradation. Although samples MB20-PLGA20-CL80 (Figure 7A–C) and MB20-PLGA50-CL50 (Figure 7D–F) retained their fibrous configuration for the whole incubation period, PAFs with the highest content of PLGA displayed significant fibre merging after 6 weeks of degradation (Figure 7H) and complete loss of mechanical integrity after 8 weeks. Both trends in mass loss and fibrous microstructure therefore reflect previous considerations regarding the higher hydrolytic reactivity of PLGA with respect to PCL [36].

The observed water uptake and degradation characteristics of PAFs were then considered to study respective MB release capability. In contrast to the complete release of MB described by sample MB20-PLGA20-CL80 within 3 h (Figure 8A), fibres with increased PLGA content revealed a significantly increased retention of MB, whereby samples MB20-PLGA80-CL20 and MB20-PLGA50-CL50 released 95 wt.% of MB only after 3 weeks (Figure 8B).

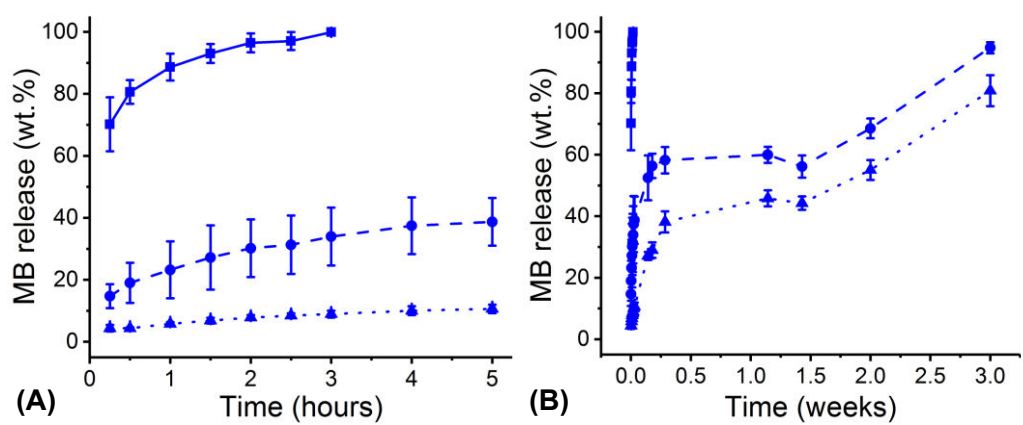

**Figure 8.** MB release profiles described by PAFs during incubation in PBS (pH 7.4, 37 °C) for 5 h (**A**) and 3 weeks (**B**). (■): MB20-PLGA20-CL80. (●): MB20-PLGA50-CL50. (▲): MB20-PLGA80-CL20. Lines are guidelines to the eye. Results are reported as Mean ± SD (n = 3).

Selected PLGA-rich fibres also indicated a slower release with respect to Indocyanine green-encapsulated PLA fibres [33], whereby ~65 wt.% release was observed within 1 week in PBS despite only 1 wt.% mass loss being recorded. In contrast, an even slower release than the one reported in this study was reported with curcumin-encapsulated polyurethane fibres with low hydrolytic degradability (≤ 5 wt.%, 2.5 weeks) [24], a result that was attributed to the hydrophobic interactions between curcumin and the fibre-forming polymer. The aforementioned release profiles therefore suggest a diffusion- rather than degradation-led mechanism of release, whereby the PLGA-rich fibres displaying ≤ 40 wt.% mass loss over three weeks described the highest retention of MB, while significant burst release was measured with the PCL- rich samples MB20-PLGA20-CL80 displaying just 5 wt.% mass loss (Figure 6B). The most likely explanation for this observation is attributed to the role played by the PLGA in mediating secondary interactions with MB, again supported by previous DSC results (Figure 3D–F, Table 2), so that diffusion of MB out of the fibres is hindered in fibres with increased PLGA content. Conversely, an excess of the PCL phase in the PAFs is demonstrated to ensure superior hydrolytic stability (Figure 6B) and wet state retained fibrous microstructure (Figure 7), in agreement with the semicrystalline morphology and hydrophobicity of this polymer.

In the following, the cytotoxicity and antibacterial testing in vitro will be presented, whereby PAFs made with equivalent PLGA:PCL weight fraction will be selected due to their controlled release and long-lasting MB retention capabilities in aqueous environment.

3.5. Contact and Extract Cytotoxicity Tests

The morphology of cells cultured on to PAFs and MB-free controls was inspected via SEM after 1- and 7-day cell culture, whereby either an initial 60-min light exposure or 60-min dark incubation (dark controls) was applied to cells. Clear cellular proliferation with evidence of ECM deposition was observed with MB-free fibres PLGA50-CL50, with cell spreading depicted already after 24 h-culture (Figure 9A and Figure S2A, Supplementary Information). At day 7, the surface of both the light-treated sample (Figure 10A) and dark control (Figure S3A, Supplementary Information) appeared completely coated by cells, confirming the high cellular tolerability of selected MB-free fibres, in line with previous reports on the biocompatibility of polyester fibres in vitro [33,37,50,57].

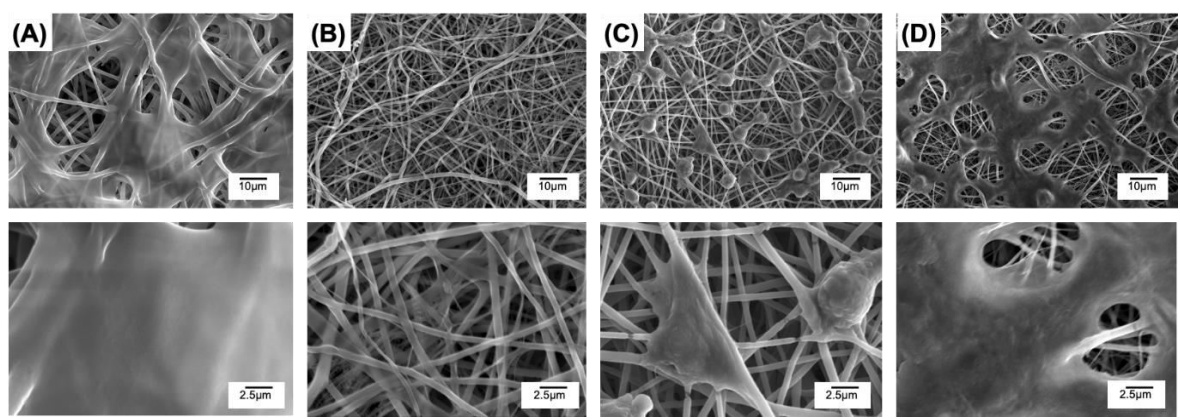

**Figure 9.** SEM of electrospun fibres cultured with L929 cells for 24 h. Prior to 24 h-culture, L929 cells were seeded on cell culture medium-equilibrated fibres and exposed to 60-min light irradiation. (**A**) PLGA50-CL50; (**B**) MB20-PLGA50-CL50; (**C**) MB10-PLGA50-CL50; (**D**) MB2-PLGA50-CL50.

Other than the electrospun controls, an indirect relationship became apparent when investigating the morphology of L929 cells cultured with fibres containing varied content of MB. No cellular attachment was visualised on sample MB20-PLGA50-CL50 electrospun with the highest concentration of MB after both 1 (Figure 9B and Figure S2B, Supplementary Information) and 7 days (Figure 10B and Figure S3B, Supplementary Information). Other than the poor cell tolerance, a fully retained fibrous microstructure was observed, in line with previous degradation results (Figures 6B and 7), suggesting that the initial 60-min light exposure did not induce any significant fibre alteration. The absence of cell attachment on these PAFs can be rationalised considering respective release profiles of MB (Figure 8) and cytotoxicity data obtained with varied MB-supplemented solutions (Figure 2A,D). An averaged MB release of either ~23 wt.% (32 µg released in 500 µL of cell culture medium (see section 2.10), i.e., ~6 μg in 100 μL) or 53 wt.% (74 μg released in 500 μL of cell culture medium (see section 2.10), i.e., ~15 μg in 100 μL) was described by aforementioned samples after 1 and 7 days, respectively. These MB concentrations were found to induce significant L929 cytotoxicity either with or without light irradiation.

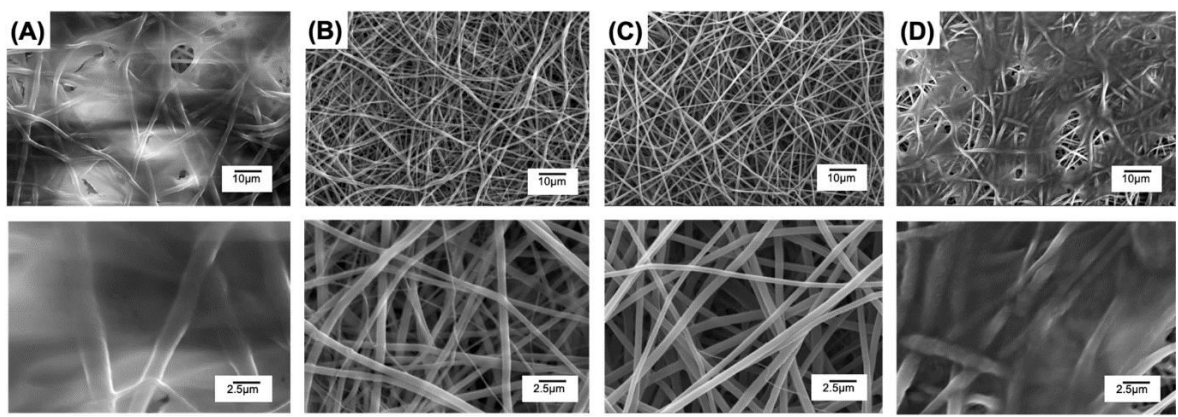

**Figure 10.** SEM of electrospun fibres cultured with L929 cells for 7 days. Prior to 7-day culture, L929 cells were seeded on cell culture medium-equilibrated fibres and exposed to 60-min light irradiation. (**A**): PLGA50-CL50; (**B**): MB20-PLGA50-CL50; (**C**): MB10-PLGA50-CL50; (**D**): MB2-PLGA50-CL50.

When sample MB10-PLGA50-CL50 with halved MB content (20 mg, $2.2 \cdot 10^{-7}$ moles MB) was tested, cell attachment was visualised after 1-day culture with both light-treated samples (Figure 9C) and corresponding dark controls (Figure S2C, Supplementary Information), even if decreased cell density and round cells were detected in comparison to the MB-free samples (Figure 9A). Extending culture from 1 to 7 days indicated significant toxic response, whereby no cell was found on the electrospun surface (Figure 10C and Figure S3C, Supplementary Information), indicating that toxic levels of MB were reached at this time point, as supported by previous release profiles (Figure 8).

It was only when sample MB2-PLGA50-CL50 ($4.4 \times 10^{-8}$ moles MB) was tested that significant cellular attachment was accomplished at both cell culture time points. Whilst round cells were detected in dark controls at day 1 (Figure S2D, Supplementary Information), this effect was somewhat less visible with light-irradiated fibres (Figure 9D), supporting the beneficial effect of light on cell viability [58–60]. After 7 days of culture, cell growth was spread over the sample surface, whereby a layer of ECM appeared to have deposited and round cells were no longer present (Figures 10D and S3D). These SEM images therefore indicate high cell tolerance of these fibres, a result that agrees with the low concentration (~0.6 and ~1.5 µg·100 µL$^{-1}$) of MB expected to be released at this time point (Figure 8). Unlike previous reports demonstrating the lack of dark cytotoxicity in vitro [24,33,35], these data confirm high cell viability even following light irradiation.

The cell tolerance of PAFs with decreased MB content were further confirmed by quantitative ATP assays on L929 cells cultured with 0–24-h fibre extracts (Figure 11). The extracts of sample MB2-PLGA50-CL50 displayed the highest degree of cell viability, whereby an averaged cell killing of 35% and 4% was measured following 60-min light activation and 60-min dark incubation, respectively. Although cell tolerance (≤ 20% ATP reduction) was also revealed in dark with the extracts of the other two fibre groups, significant cytotoxic effects (≥ 70% ATP reduction) were measured in both cases following activation with light, providing first indication of fibre photodynamic activity.

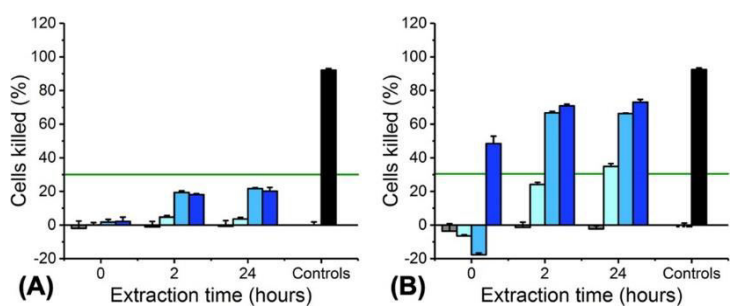

**Figure 11.** ATP toxicity assay on L929 cells cultured with 0–24-h extracts of fibres MBX-PLGA50-CL50. (**A**) 60-min dark incubation. (**B**) 60-min light irradiation. (■): MB-free control PLGA50-CL50; (■): MB20-PLGA50-CL50; (■): MB10-PLGA50-CL50; (■): MB2-PLGA50-CL50; (□): PBS negative control; (■): OPDEGE positive control. (—): 30% reduction in cell viability, above which cytotoxic response is observed (ISO 10993). Results reported as Mean ± SD (n = 3).

*3.6. Antibacterial Photodynamic Capability of PAFs*

Once the cytotoxicity of PAFs was investigated, the attention moved to the quantification of the antibacterial photodynamic capability of PAFs. Fibres electrospun with varied MB content were initially cultured in contact with agar plates inoculated with either *E. coli* or S*. mutans*, so that the zone of inhibition (ZOI) following light irradiation (for either 30 or 60 min) and dark incubation was measured (Figure 12). Incubation of MB-free fibre controls revealed no detectable effect on both aforementioned bacteria, confirming the lack of antibacterial properties in pristine polyester fibres. Conversely, significant ZOIs were measured with either 30-min (Ø = 11.7 ± 0.3–12.7 ± 0.82 mm) or 60-min (Ø = 15.6 ± 2.1–18.7 ± 1.0 mm) light activation in *E. coli* inoculated agar plates, whereby variation of the MB content encapsulated in the fibres proved to play an insignificant effect (Figure 12A).

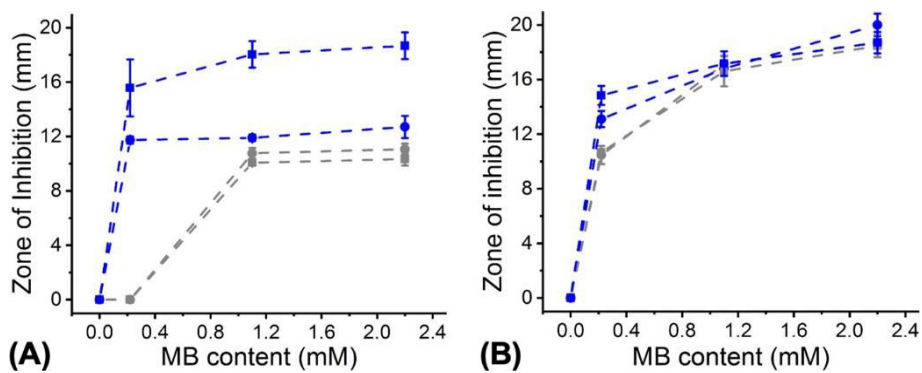

**Figure 12.** Zone of inhibition of *E. coli* (**A**) and *S. mutans* (**B**) following testing with PAFs electrospun with varied MB content. (■): 60-min light irradiation; (■) 60-min dark incubation. (●): 30-min light irradiation; (●): 30-min dark incubation. Results are reported as mean ± SD (n = 3). Lines are guideline to the eye.

In comparison, lower but still detectable ZOIs (Ø = 10.1 ± 0.1–11.1 ± 0.4 mm) were measured when PAFs electrospun with increased content of MB were incubated in dark with *E coli*-inoculated agar plates, suggesting a dosage-dependent bactericidal effect with inert MB, as observed with the PS-supplemented solutions (Figure 2B,C). Other than *E. coli*, significant antibacterial effects were also observed against *S. mutans* (Figure 12B), whereby variation in fibre MB dosage, rather than the irradiation time, seemed to play a significant role. ZOIs in the range of 14.8 ± 0.7–

18.7 ± 0.9 mm were measured on the 60-min light-treated samples, which proved to be similar to the one recorded following dark incubation (Ø = 10.7 ± 0.7–18.5 ± 1.0 mm). This observation is in agreement with previous photodynamic data, whereby similar viability reduction was expressed by *S. mutans* after 60-min incubation with MB-supplemented solutions in dark (Figure 2C) and under light irradiation (Figure 2F).

To confirm that the photodynamic effect was governed by the release of MB out of the PAFs, 0–24-h extracts of fibre MB2-PLGA50-CL50 were tested and bacteria counts measured (Figure 13).

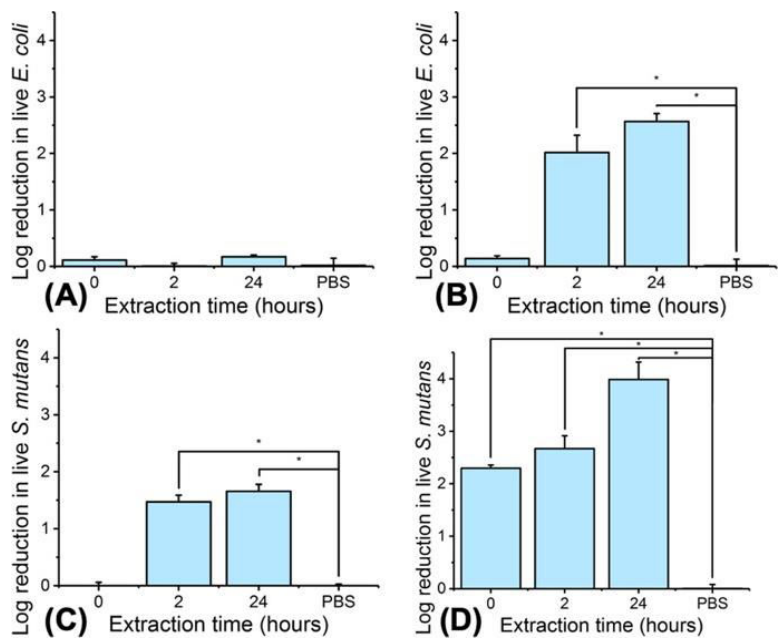

**Figure 13.** Agar plate testing of E.coli (**A–B**) and S. mutans (**C–D**) with 0–24-h extract of fibres MB2-PLGA50-CL50. (A,C) 60-min incubation in dark. (B,D) 60-min light irradiation. Results reported as mean ± SD (n = 3). '*' denotes significantly different means ($p < 0.05$, t-test).

In contrast to the results obtained in dark (≤ 0.1 log reduction) and with the PBS negative control, a significant reduction (2–2.5 log) in *E. coli* viability was again recorded following light treatment, whereby an extraction time of at least two hours was needed to enable detectable effects on *E. coli*. When the same test was carried out with *S. mutans*, a significant reduction (> 1 log) was still observed following dark incubation with both 2 and 24-h fibre extract, a result that is again in agreement with the decreased tolerance of this bacteria to native MB (Figure 2C). These values were further increased (> 2 log reduction) when light was applied, whereby even the 0-h extract was proven to play a significant antibacterial effect. Overall these data confirm the key role played by the release capability of PAFs in triggering photodynamic effect on-demand and ensuring bacterial tolerance when no light exposure is applied. In this regard, the release profile of the PS should also be customised according to the target bacteria, given aforementioned difference in photodynamic susceptibility according to the bacterial structure.

## 4. Conclusions

Electrospun PAFs with varying PLGA:PCL weight fraction and MB content were studied to accomplish retained wet-state fibrous microstructure, controlled degradability and controlled PS release profile, as a means to deliver long-lasting fibres with integrated cellular tolerance and antibacterial photodynamic effect. Preliminary in vitro studies with MB- and ER-supplemented solutions successfully indicated the superior uptake of MB by L929 cells, *E. coli* and *S. mutans*, as well as an increased photodynamic effect of MB, which informed the formulation of PAFs. MB-encapsulated electrospun fibres were successfully realised with ~100 wt.% encapsulation efficiency, semicrystalline polymer morphology and MB-PLGA secondary interactions, generating complete (95 wt.%) and controlled release of MB over 3 weeks (37 °C, pH 7.4), as well as retained fibrous microstructure following up to 8-week incubation in water. In vitro, PAFs with increased PLGA content promoted attachment and proliferation of L929 cells over a 7-day culture with or without an initial 60-min light exposure, successfully indicating cell tolerance of both the fibres and released MB. Most importantly, up to 2.5 and 4 log viability reduction and up to ~19 mm zone of inhibition were measured when the same material was tested with

*E. coli* and *S. mutans*, respectively, confirming the antibacterial photodynamic capability of resulting fibres with selected bacteria. These results therefore provide new perspectives on the integration of antibacterial PDT in fibrous materials, aiming to accomplish durable and selective antibacterial activity with minimal impact on cell viability. In light of their fibroblast-friendly surface, the presented PAFs could therefore be used to support wound healing and offer a platform to treat antibiotic-resistant infections limiting risks of detrimental impact on host tissues and cells. Ultimately, the FDA-approved status of both aforementioned fibre-forming polymers and MB is key from a translation viewpoint, as it provides an underpinning framework aiming to secure regulatory compliance for respective MB-encapsulated fibres.

**Funding**

This research was funded by the EPSRC Centre for Doctoral Training in Tissue Engineering and Regenerative Medicine, grant number EP/L014823/1, and Neotherix Limited.

**Acknowledgments**

The authors also wish to thank Jackie Hudson and Sarah Myers for technical assistance with SEM and cell culture, respectively.

**Conflicts of Interest**

The authors declare the following competing financial interest(s): Michael J. Raxworthy is CEO of Neotherix Limited.

## Supplementary Information

### Light source and intensity measurements

A 6000-lumen work light (50W, 135 lumen/W, 2800-3200 warm light) was selected as a light source model. A hand-held optical meter (ILT2400, International Light Technologies) was used with a laser line filter (ThorLabs, Inc.) centred at 670±2 nm to determine the light intensity in the spectral regions of MB peak absorbance. The light intensity (mW·cm$^{-2}$) was measured with and without the filter in 9 distinct locations of the lamp and values averaged. The measurement was repeated three times.

### Encapsulation efficiency of PAFs

Discs (Ø 1 cm) of PAFs were individually weighed on an analytical balance prior to 48- incubation in HFIP (5 ml) to enable complete sample dissolution. A standard calibration curve was built via UV-Vis spectrophotometry using MB solutions in HFIP covering the range of MB concentration expected in PAFs. The encapsulation efficiency (EE) of MB in PAFs was calculated according to Equation S1:

$$EE = \frac{m_d}{m_e} \times 100,$$  Equation (S1)

whereby $m_d$ and $m_e$ are the determined and expected weights of MB in the electrospun samples, respectively.

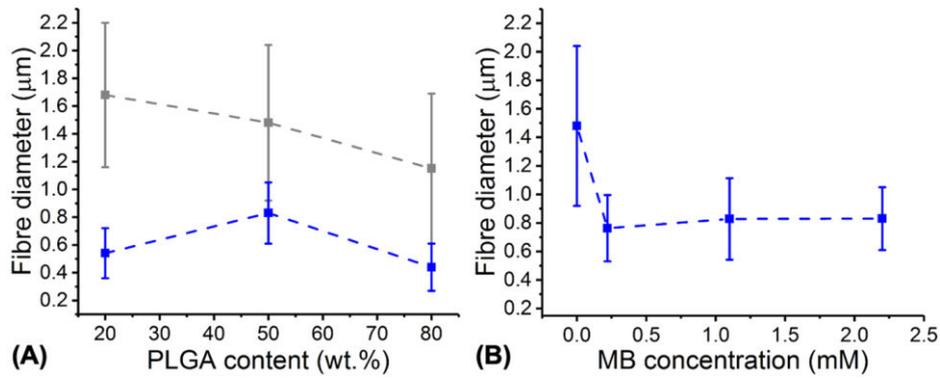

**Figure S1.** (A): Variation of fibre diameter in PAFs (■) and MB-free fibre controls (■) made from electrospinning solutions containing varied PLGA content and constant MB concentration (2.2 mM). (B) Variation of fibre diameter in PAFs made from electrospinning solutions containing constant PLGA content (50 wt.%) and varied MB concentration.

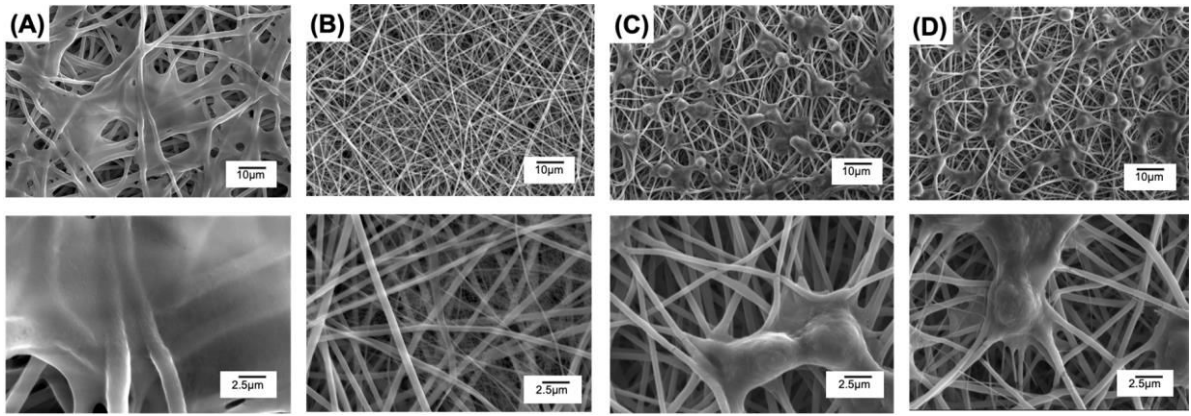

**Figure S2.** SEM of electrospun fibres cultured with L929 cells for 24 hours. Prior to 24-hour culture, L929 cells were seeded on cell culture medium-equilibrated fibres and exposed to 60-min light irradiation in dark. (A): PLGA50-CL50; (B): MB20-PLGA50-CL50; (C): MB10-PLGA50-CL50; (D): MB2- PLGA50-CL50.

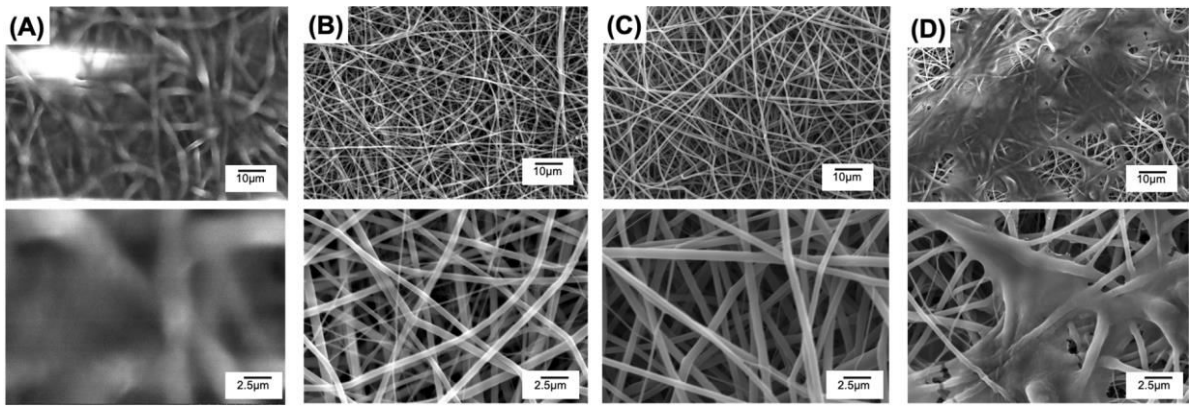

**Figure S3.** SEM of electrospun fibres cultured with L929 cells for 7 days. Prior to 7-day culture, L929 cells were seeded on cell culture medium-equilibrated fibres and exposed to 60-min light irradiation in dark. (A): PLGA50-CL50; (B): MB20-PLGA50-CL50; (C): MB10-PLGA50-CL50; (D): MB2-PLGA50-CL50.